\documentclass[%
 reprint, 
nofootinbib,
 amsmath,amssymb,
 aps, physrev,
prc,
longbibliography
]{revtex4-1}
\usepackage{graphicx}
\usepackage{dcolumn}
\usepackage{bm}
\usepackage{hyperref}
\usepackage[capitalise]{cleveref}

\usepackage{color}

\begin{document}


\title{\textbf{Energy loss baseline for light hadrons in oxygen-oxygen collisions at $\sqrt{s_\mathrm{NN}}=5.36\,\text{TeV}$} 
}%

\author{Aleksas Mazeliauskas}
\email{a.mazeliauskas@thphys.uni-heidelberg.de}
\affiliation{Institute for Theoretical Physics, University of Heidelberg, 69120 Heidelberg, Germany}

\date{\today}

\begin{abstract}
I present predictions for inclusive charged hadron spectra in minimum-bias proton-proton and oxygen-oxygen collisions at a centre-of-mass energy of $\sqrt{s_\mathrm{NN}} = 5.36\,\text{TeV}$, assuming no final-state interactions. Using next-to-leading order perturbative QCD matrix elements, along with state-of-the-art (nuclear) parton distribution and fragmentation functions, I establish a baseline for the nuclear modification factor $R^h_\text{AA}$ in oxygen-oxygen collisions in the absence of quenching. Theoretical uncertainties in this baseline are found to be substantial for transverse momenta below $20\,\text{GeV}$. In the intermediate range $20\,\text{GeV} \lesssim p_T^h \lesssim 70\,\text{GeV}$, these uncertainties are significantly reduced to approximately 5\%. At higher momenta ($p_T^h \gtrsim 70\,\text{GeV}$), however, predictions exhibit a marked spread due to differences between fragmentation functions, reflecting varying assumptions about isospin symmetry. Finally, I show that considering neon-neon collisions in the initial state or neutral pions in the final state does not appreciably change the nuclear modification factor.
\end{abstract}

\maketitle


\section{\label{sec:level1}Introduction}

Collisions of ultra-relativistic heavy nuclei at the Relativistic Heavy Ion Collider (RHIC) and the Large Hadron Collider (LHC) provide a unique window into a novel state of quantum matter, the Quark–Gluon Plasma (QGP), composed of deconfined quark and gluon degrees of freedom~\cite{Busza:2018rrf}.
A key signature of the formation of QGP is the suppression of high-momentum hadron and jet yields (jet quenching). This phenomenon is understood as energy loss of high-momentum partons, produced in initial hard scatterings, through multiple interactions with the medium, which leads to a depletion of the observed high-momentum yield~\cite{Cunqueiro:2021wls,Connors:2017ptx,Qin:2015srf}.
A major open question in the field is understanding how different QGP signals, such as jet quenching, evolve with the size of the colliding system~\cite{Nagle:2018nvi,Grosse-Oetringhaus:2024bwr,Noronha:2024dtq}.

Strikingly, several signatures traditionally associated with QGP formation, such as azimuthal anisotropies and strangeness enhancement, have been observed even in small systems, such as proton-proton, proton-nucleus, and peripheral nucleus-nucleus collisions~\cite{CMS:2015yux,ALICE:2015mpp,CMS:2016fnw,ALICE:2016fzo}. 
However, an unambiguous signal of partonic energy loss has not yet been established in systems with  $\lesssim 30$ participating nucleons.
Centrality-selected peripheral heavy-ion collisions (e.g., PbPb at the LHC or AuAu at RHIC) are complicated by significant uncertainties in collision-geometry modelling as well as by centrality-selection biases~\cite{Loizides:2017sqq,dEnterria:2020dwq}.
Measurements in asymmetric proton–nucleus collisions have so far provided only upper limits on possible jet-quenching effects~\cite{ALICE:2017svf,ATLAS:2022iyq,CMS:2025jbv}.
A notable exception are recent PHENIX results in deuteron–gold collisions, where a suppression of neutral pion production was observed in the highest event-activity class~\cite{PHENIX:2023dxl}; however, its interpretation in terms of a genuine final-state effect remains under debate~\cite{Perepelitsa:2024eik}.

Collisions of light ions provide symmetric systems with significant advantages and unique physics opportunities~\cite{Citron:2018lsq,Brewer:2021kiv,light_ions}. In particular, oxygen ion ($^{16}_{\phantom{1}8}\mathrm{O}$) collisions involve, on average, $\sim 10$ participating nucleons, thus bridging the system size range between proton-nucleus and peripheral nucleus-nucleus collisions~\cite{Loizides:2025ule}. In March 2021, RHIC conducted oxygen–oxygen collisions at $\sqrt{s_\mathrm{NN}} = 200\,\text{GeV}$~\cite{Liu:2022jtl}. More recently, in July 2025, the LHC carried out oxygen–oxygen and neon–neon collisions at $\sqrt{s_\mathrm{NN}} = 5.36\,\text{TeV}$, as well as proton–oxygen collisions at $\sqrt{s_\mathrm{NN}} = 9.62\,\text{TeV}$~\cite{oxygen_collisions_LHC,163rdLHCC}. A pressing open question is whether unambiguous evidence of partonic energy loss can be established in these small collision systems.

Selecting central oxygen-oxygen collisions enhances the potential medium-induced energy loss signal. However, such a selection introduces potential biases~\cite{Park:2025mbt} and, importantly, prevents the construction of an accurate theoretical baseline. Only in minimum-bias collisions can the inclusive high-momentum hadron and jet production be reliably computed using QCD factorization. The standard observable of energy loss is the nuclear modification factor $R_\text{AA}$, which is the ratio of  $p_T$ spectra in nucleus-nucleus and proton-proton collisions scaled by the number of binary collisions~\cite{Cunqueiro:2021wls,Connors:2017ptx,Qin:2015srf}. In minimum-bias collisions $R_\text{AA}$ can be written as a ratio  of $p_T$-differential cross sections (for some rapidity window $|y| < y_\text{max}$):
\begin{equation}
R_\mathrm{AA} \equiv \frac{1}{A^2}\frac{d\sigma_\mathrm{OO}/dp_T}{d\sigma_{pp}/dp_T}. \label{eq:RAA}
\end{equation}
where $A=16$ is the nuclear mass number.

In this work, I compute the perturbative QCD baseline, i.e., the expectation in the absence of medium-induced energy loss, for the nuclear modification factor $R^h_\mathrm{AA}$ of single-inclusive charged hadrons and neutral pions in minimum-bias oxygen–oxygen collisions at $\sqrt{s_\mathrm{NN}} = 5.36\,\text{TeV}$\footnote{All scripts and data required to reproduce the plots are available at \url{https://github.com/amazeliauskas/OO_baseline}}. Any experimentally observed deviation from this “no-quenching” baseline would signal final-state effects indicative of QGP formation. Since the expected energy-loss signal is small, it is essential to quantify theoretical uncertainties in the baseline of \cref{eq:RAA} with high accuracy. To this end, I perform next-to-leading-order (NLO) perturbative QCD calculations using state-of-the-art nuclear parton distribution functions (nPDFs) and hadron fragmentation functions (FFs). This work updates the previous baseline estimate of Ref.~\cite{Huss:2020dwe} with latests nPDFs, FFs and the actual LHC centre-of-mass energy for oxygen–oxygen collisions. Baseline estimates for charged hadron $R^h_\mathrm{AA}$ for RHIC energies can be found in Ref.~\cite{Belmont:2023fau}. Baseline computations for inclusive jet $R^j_\mathrm{AA}$ and semi-inclusive jet- and hadron-triggered nuclear modification factors were recently presented in Ref.~\cite{Gebhard:2024flv}.
Although a statistically significant deviation between the no-quenching baseline and experimental data would suffice to establish energy loss in oxygen–oxygen collisions, interpreting such a signal requires dedicated model predictions. These have been developed in the literature~\cite{Huss:2020whe,Liu:2021izt,Zakharov:2021uza,Ke:2022gkq,Behera:2023oxe,Xie:2022fak,vanderSchee:2025hoe,Pablos2025} and are not discussed in this paper.

The paper is structured as follows. In \cref{sec:setup}, I describe the computational framework and present a validation plot. \Cref{sec:results} contains the baseline results for the charged-hadron and neutral-pion nuclear modification factor $R_\text{AA}$. For completeness, I also provide the absolute cross-section predictions for proton–proton and oxygen–oxygen collisions separately. The conclusions are summarised in \cref{sec:discussion}. Additional results for different rapidity selections are shown in \cref{app:rapidity}, while \cref{app:isospin} discusses the role of isospin symmetry in fragmentation functions.

\section{Computational setup\label{sec:setup}}

\subsection{Inclusive hadron cross-section}

According to QCD factorisation, the perturbative single-inclusive hadronic cross section  is given by the convolution of parton distribution functions, hard partonic cross section and fragmentation functions~\cite{Collins:1989gx,Metz:2016swz}
\begin{equation}
    E^h\frac{d\sigma^{AB\to h}}{d^3p^h}= f^{A}_{i}\otimes  f^B_j  \otimes \hat{\sigma}^{ij\to l}  \otimes  D^h_{l}.\label{eq:sgh}
\end{equation}
Here $f^{A}_{i}$ and $f^{B}_{j}$ are parton distribution functions (PDFs) for nucleus $A$ and $B$ and partons $i$ and $j$ at factorization scale $\mu_F$. In this paper, I consider only symmetric collision systems, i.e., $A=B=p$ or $A=B={}^{16}_{\phantom{1}8}\mathrm{O}$.
$\hat{\sigma}^{ij\to l}$ is the partonic cross-section $ij\to l$, which is computed at leading (LO) and next-to-leading (NLO) order at renormalization scale $\mu_R$.
Finally, $D^h_{l}$ is the parton $l$ fragmentation function (FF) into a hadron species $h$ at fragmentation scale $\mu_{FF}$. I consider two cases: fragmentation to charged hadrons and fragmentation to neutral pions, i.e., $h=\pi^0$.

I evaluated \cref{eq:sgh} using \texttt{INCNLO} program~\cite{Aversa:1988vb}\footnote{\url{http://lapth.cnrs.fr/PHOX_FAMILY/readme_inc.html}} modified to use LHAPDF grids~\cite{Buckley:2014ana} for parton distribution and fragmentation functions. Modifications were checked against an independent LO implementation of the hadron cross-section formula~\cite{Huss:2020dwe}. \texttt{INCNLO}  computes the hadron cross section differentially in transverse momentum $p_T$, but averaged in the azimuth and rapidity $|y|<y_\text{max}$, i.e., $d\sigma^h/(2\pi p_T dp_T dy)$.
The factorisation, renormalisation, and fragmentation scales are chosen to be equal to hadron transverse momentum $\mu_F=\mu_R=\mu_{FF}=p_T^h$. 

To estimate uncertainties from missing higher order terms, I vary $\mu_F,\mu_R$ and $\mu_{FF}$ by a factor $2$ up and down with a condition that no two scales are varied up and down at the same time (15-point scale uncertainty prescription). Although scale uncertainties do not have probabilistic interpretation~\cite{Duhr:2021mfd}, the nested LO and NLO scale uncertainties are conventionally used to assess the perturbative convergence of the observable.

To estimate uncertainties from (nuclear) parton distribution functions, I use three sets of proton and oxygen PDFs:  EPPS21~\cite{Eskola:2021nhw}, which uses CT18ANLO~\cite{Hou:2019efy} as a proton PDF baseline, nNNPDF3.0~\cite{AbdulKhalek:2022fyi}, which is matched to NNPDF3.1~\cite{NNPDF:2017mvq} proton PDF, and TUJU21~\cite{Helenius:2021tof}, which simultaneously determines nucleus and proton PDFs. All three nPDF determinations include experimental data from measurements in $p$Pb collisions at the LHC:  $W^\pm$ and $Z$ boson production (TUJU21, EPPS21, nNNPDF3.0); dijet and $D$-meson data (EPPS21, nNNPDF3.0); isolated photon measurements (nNNPDF3.0).
Each nPDF extraction is accompanied by a large set of PDF variations, which can be used to determine the confidence intervals for the computed cross section. For each PDF in an error set, I repeat the cross-section computation and use the built-in LHAPDF functionality to compute 68\% confidence intervals. Note that no data with oxygen nucleus is included in the fits, and oxygen nPDFs are determined by interpolation in nuclear mass number $A$. Comparison between different collaborations shows how robust this extraction is with respect to changes in the input data, fitting techniques, and nPDF parametrisation.

To estimate uncertainties from the choice of fragmentation functions, I used three choices of fragmentation functions: Binnewies-Kniehl-Kramer (BKK)~\cite{Binnewies:1994ju} FFs, which is a built-in option in \texttt{INCNLO}, NNFF1.1h\cite{Bertone:2018ecm}\footnote{The initial NNFF1.1h release contained an error that is corrected in the erratum. I used corrected nPDF grids available at \url{https://nnpdf.mi.infn.it/nnff1-1h/}} and NPC23~\cite{Gao:2024dbv} FFs. BKK fits charged pion and kaon FFs to $\sqrt{s}=29\,\text{GeV}$ $e^+e^-$ annihilation data. The charged hadron FFs are formed as a sum of charged pion and kaon FFs. Pion contribution is increased by a factor $1.16$ to mimic (anti)proton contribution. NNFF1.1h charged hadron FFs are obtained from a direct fit to charged hadron data from a large set of single-inclusive $e^+e^-$ annihilation (SIA) and $pp$ collisions. Notably, NNFF1.1h includes LHC charged hadron data at centre of mass energies $\sqrt{s}=0.9,2.76$ and $7\,\text{TeV}$. NPC23 performs
global analysis of a wide range of precision measurements from the LHC, as well as data from electron-positron collisions and semi-inclusive deep inelastic scatterings. NPC23 includes both identified hadron ($\pi^\pm, K^\pm, p(\bar p)$) and charged hadron data at LHC up to $\sqrt{s}=13\,\text{TeV}$. Charged hadron FFs are a sum of fitted charged pion, kaon, and (anti)proton FFs.
We also use neutral pion FFs constructed from NPC23 charged pion FFs using isospin symmetry~\cite{Gao:2025bko}\footnote{FF grids for neutral pions were obtained from the authors by private communication.}. NNFF1.1h and NPC23 provide FF error sets, which are used to compute 68\% confidence intervals. BKK FFs do not include uncertainty estimates, and only scale uncertainties are shown.

In the ratio of cross sections \cref{eq:RAA}, one can expect a significant cancellation between common sources of theoretical uncertainties in proton and oxygen cross sections. Therefore, the scale uncertainty of $R_\text{AA}$ is determined by the simultaneous variation of $\mu_F,\mu_{R}$ and $\mu_{FF}$ in the numerator and denominator. Cross sections for different members of the oxygen nPDF error set are also matched to the corresponding error set members in the proton baseline, maximising the uncertainty cancellation. For NNFF1.1h and NPC23 FFs, uncertainties in $R_\text{AA}$ are computed by taking the cross-section ratio for each member in the error set and then applying the corresponding uncertainty prescription.

\subsection{Validation}

\begin{figure}
    \centering
    \includegraphics[width=\linewidth]{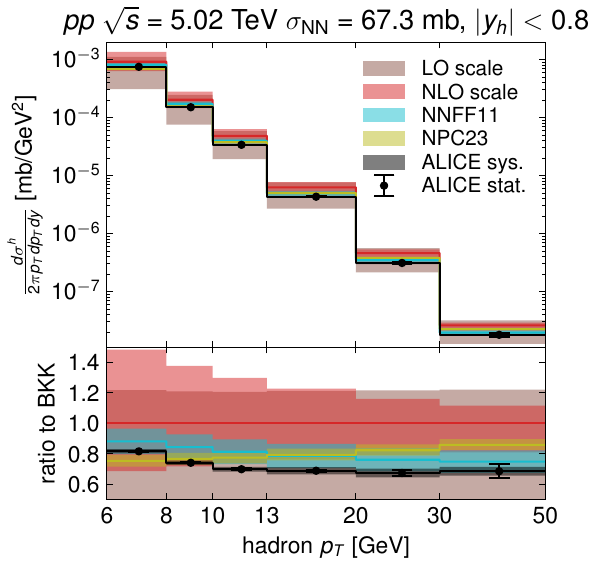}
    \caption{pQCD computation of differential charged hadron cross section in minimum-bias $pp$ collisions at $\sqrt{s}=5.02\,\text{TeV}$ with CT18ANLO PDF. The lower panel shows the ratio to NLO BKK FF results.  The brown and red colour bands represent LO and NLO scale variation for BKK FFs. Cyan and olive bands indicate 68\% confidence intervals for NNFF1.1h and NPC23 FFs. Black bands and error points show systematic and statistical uncertainties of ALICE measured differential charge particle yield~\cite{ALICE:2018vuu}, which is converted to cross section using $pp$ inelastic cross section of $\sigma_{pp}^\text{NN}=67.3\,\text{mb}$~\cite{dEnterria:2020dwq}\label{fig:5020spectra}}
\end{figure}

To validate numerical setup,  I computed inclusive hadron cross section in proton-proton collisions at $\sqrt{s}=5.02\,\text{TeV}$, $6\,\text{GeV}<p_T^h<50\,\text{GeV}$ and $|y|<0.8$ using CT18ANLO PDF (central value) with BKK, NNFF1.1h and NPC23 FFs (c.f. Ref.~\cite{Brewer:2021tyv}). The results are shown in \cref{fig:5020spectra} and compared to experimental measurements by ALICE~\cite{ALICE:2018vuu}. The $p_T$-binned cross section was obtained using a three-point Gauss–Legendre quadrature to evaluate the momentum integrals. Note that  ALICE reports the averaged hadron yield in $pp$ collisions, which can be converted to the cross section by scaling it with inelastic $pp$ cross section $\sigma_\text{NN}=67.3\pm 1.2\,\text{mb}$~\cite{dEnterria:2020dwq}. The black bands and black error points correspond to experimental systematic and statistical uncertainties, respectively (uncertainty in $\sigma_\text{NN}$ is neglected). The lower panel shows the ratio to NLO BKK FF results.

The brown and red colour bands indicate the LO and NLO scale uncertainties computed with CT18ANLO PDFs and BKK FFs. The NLO band ranges from $\sim 10\%$ in the highest momentum bin to almost 50\% in the lowest momentum bin. The NLO central value is within the (large) LO scale band. Experimental data points are below NLO BKK expectations, even taking into account the scale uncertainty (see also Refs.~\cite {Brewer:2021tyv,dEnterria:2013sgr}). However, the ratio is approximately constant, which indicates that a suitable choice of fragmentation scale can better reproduce the spectra, see Ref.~\cite{Borsa:2021ran}.

In cyan and olive bands, I show 68\% confidence intervals computed with NNFF1.1h and NPC23 charged hadron FFs. NNFF1.1h results are only slightly shifted above the data. NPC23 results have a smaller uncertainty band, but the momentum dependence is different from the one observed in data. Taking into account the scale uncertainty, NNFF1.1h and NPC23 results are in good agreement with experimental data, which is not surprising, since these FFs are fitted to LHC proton-proton data. Finally, I note that proton PDFs are much better constrained than FFs, and the corresponding (much smaller) uncertainty is not shown in \cref{fig:5020spectra}. See the predictions at $\sqrt{s}=5.36\,\text{TeV}$ in \cref{sec:spectra} and Ref.~\cite{Brewer:2021tyv} for proton PDF uncertainties in the spectrum.

\section{Results\label{sec:results}}
\subsection{Nuclear modification factor}

\begin{figure}
    \centering
    \includegraphics[width=\linewidth]{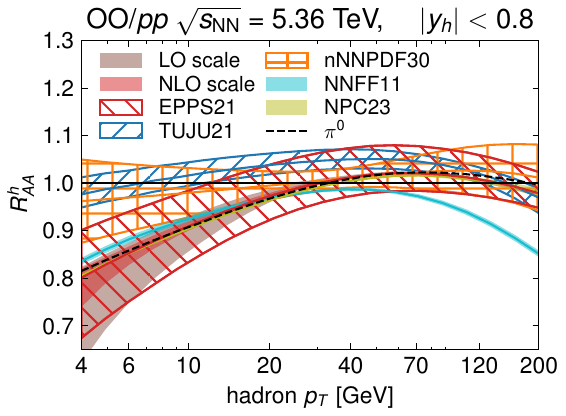}
    \caption{
    pQCD calculation of the charged-hadron nuclear modification factor baseline in oxygen–oxygen collisions at $\sqrt{s_\text{NN}}=5.36\,\text{TeV}$. Hatched bands denote the 68\% confidence intervals of EPPS21, TUJU21, and nNNPDF3.0 nPDFs. Brown and red bands indicate LO and NLO scale variations with BKK FFs. Cyan and olive bands show the 68\% confidence intervals for NNFF1.1h and NPC23 FFs. The black dashed line represents the central prediction for the neutral-pion nuclear modification factor obtained with NPC23 FFs.}
    \label{fig:RAA}
\end{figure}

In \cref{fig:RAA}, I present the no-quenching baseline results for the charged-hadron nuclear modification factor $R^h_\text{AA}$ in oxygen–oxygen collisions at $\sqrt{s}=5.36\,\text{TeV}$, within the kinematic range $|y|<0.8$ and $4\,\text{GeV} < p_T < 200\,\text{GeV}$. Additional results for other rapidity intervals are provided in \cref{app:rapidity}.

The brown and red shaded bands indicate the LO and NLO scale uncertainties, computed with EPPS21 nPDFs and BKK FFs. At $p_T \gtrsim 20\,\text{GeV}$, these uncertainties cancel efficiently in the ratio, and the NLO band becomes comparable to the line thickness. At lower $p_T$, scale uncertainties grow more noticeable, yet the NLO band remains fully nested within the LO band, reflecting good perturbative convergence.

The hatched red, blue, and orange bands correspond to the 68\% confidence intervals of EPPS21, TUJU21, and nNNPDF3.0 nPDFs (all with BKK FFs). For $p_T \gtrsim 20\,\text{GeV}$, the three nPDFs are consistent, with overlapping uncertainty bands. In the range $20\,\text{GeV} \lesssim p_T \lesssim 100\,\text{GeV}$, EPPS21 gives the widest uncertainty band ($\sim 5\%$), while for $p_T \gtrsim 100\,\text{GeV}$, nNNPDF3.0 becomes slightly broader. TUJU21 shows the narrowest uncertainties, though this reflects its limited dataset and restrictive parametrisation~\cite{Helenius:2021tof}. Interestingly, the central values of all three nPDFs yield $R^h_\text{AA} > 1$ in the interval $30\,\text{GeV} \lesssim p_T \lesssim 120\,\text{GeV}$, though still consistent with unity within EPPS21 and nNNPDF3.0 uncertainties.

\begin{figure}
    \centering
    \includegraphics[width=\linewidth]{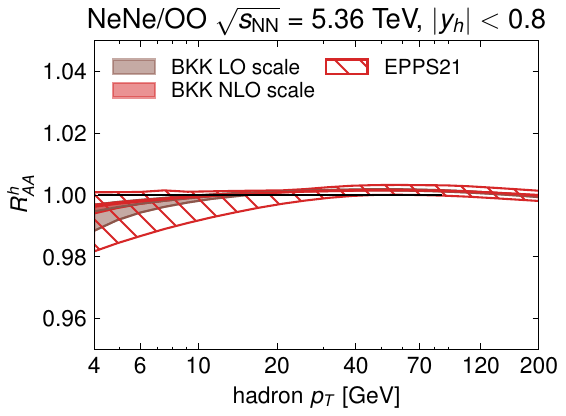}
    \caption{QCD calculation of the charged-hadron cross-section ratio in neon–neon and oxygen–oxygen collisions at $\sqrt{s_\text{NN}}=5.36\,\text{TeV}$ for $|y|<0.8$. Coloured bands indicate scale variations, while the hatched band shows the 68\% confidence interval of EPPS21 nPDFs.}
    \label{fig:NeO}
\end{figure}

At lower momenta ($p_T \lesssim 20\,\text{GeV}$), nPDF uncertainties increase, and differences between the extractions become more pronounced. TUJU21 and nNNPDF3.0 remain centred near unity, whereas the EPPS21 band drops below unity for $p_T \lesssim 10\,\text{GeV}$. Thus, experimental observations of $R^h_\text{AA} < 1$ in this region would not unambiguously signal medium-induced energy loss.

Next, I examine the sensitivity to fragmentation functions (FFs). The cyan and olive bands show the 68\% confidence intervals for NNFF1.1h and NPC23 charged-hadron FFs. For both uncertainty cancellation is excellent with bands barely wider than the line thickness across the full $p_T$ range. Results with NPC23 closely match those with BKK FFs, whereas NNFF1.1h exhibits striking deviations: at $p_T \gtrsim 70\,\text{GeV}$, $R^h_\text{AA}$ falls well below unity. As discussed in \cref{app:isospin}, this originates from strong isospin breaking in NNFF1.1h, where down-quark fragmentation is significantly suppressed. Since oxygen nuclei contain a larger fraction of down quarks, the hadron yield is reduced at high $p_T$, where quark fragmentation dominates~\cite{Huss:2020whe}. Such strong isospin breaking is unexpected, given that charged hadrons are dominated by pions. Nevertheless, in the intermediate region ($20\,\text{GeV} \lesssim p_T \lesssim 70\,\text{GeV}$), differences between FF sets remain below 5\%.

The black dashed line in \cref{fig:RAA} shows the central prediction for the neutral-pion $R_\text{AA}$ obtained with NPC23 FFs. It closely follows the charged-hadron $R^h_\text{AA}$ with the same FF set, as expected, since charged-hadron production is pion-dominated and the neutral-pion FFs are constructed from charged-pion FFs.

\begin{figure*}
\centering
    \includegraphics[width=0.45\linewidth]{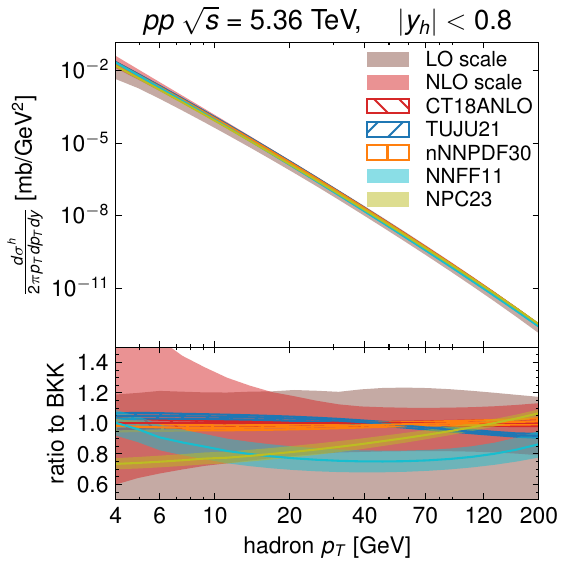}
    \includegraphics[width=0.45\linewidth]{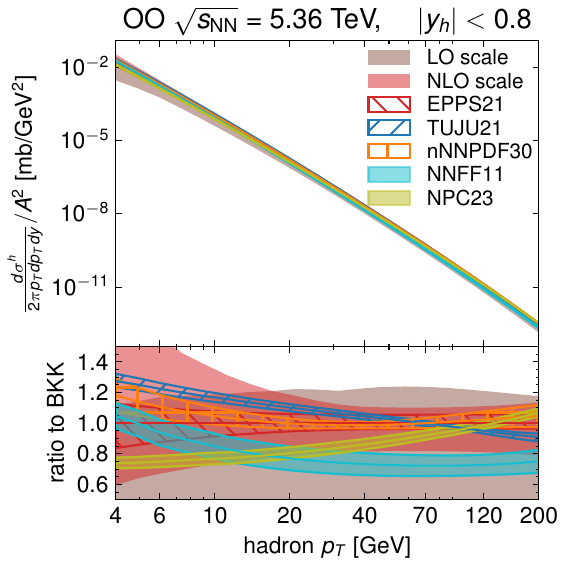}
    \caption{pQCD computation of differential charged hadron cross section in minimum-bias (left) $pp$ and (right) oxygen-oxygen collisions at $\sqrt{s_\text{NN}}=5.36\,\text{TeV}$. The brown and red bands show scale variation uncertainty. The hatched bands represent 68\% confidence intervals of (left) CT18ANLO, TUJU21 and nNNPDF3.0 PDFs and (right) EPPS21, TUJU21 and nNNPDF3.0 nPDFs. The cyan and olive bands show 68\% confidence intervals of NNFF1.1h and NPC23 FFs. Note that the oxygen-oxygen cross section does not include final state modification and is normalised by $A^2$. }
    \label{fig:spectra}
\end{figure*}

Finally, \cref{fig:NeO} presents the ratio of differential hadron cross sections (without final-state interactions) in neon–neon and oxygen–oxygen collisions at $\sqrt{s_\text{NN}}=5.36\,\text{TeV}$\footnote{I downloaded nPDF grids for $^{20}_{10}$Ne from  \url{https://research.hip.fi/qcdtheory/nuclear-pdfs/epps21/}.}. The ratio agrees with unity to within 1\% over most of the momentum range. This reflects the fact that the nuclear mass number dependence is poorly constrained in current nPDFs, leading to strongly correlated and nearly identical modifications for oxygen and neon. Consequently, the nuclear modification factor baseline for neon–neon collisions can be taken as essentially identical to that for oxygen–oxygen within current knowledge of nPDFs.

\subsection{Differential hadron cross section\label{sec:spectra}}

For completeness, in \cref{fig:spectra} I present pQCD calculations of the absolute $p_T$-differential hadron cross section in minimum-bias (left) $pp$ and (right) oxygen–oxygen collisions at $\sqrt{s}=5.36\,\text{TeV}$ with $|y|<0.8$. The OO spectrum does not include final-state modifications and is normalised by the square of the nuclear mass number, $A^2$. The lower panels display the ratios to the central NLO prediction obtained with BKK FFs, using (left) CT18ANLO PDFs and (right) EPPS21 nPDFs.

The brown and red bands indicate the LO and NLO scale uncertainties. At LO, the uncertainty remains large across the entire $p_T$ range. At NLO, the band is largely contained within the LO one, except for $p_T^h<20\,\text{GeV}$. At high momenta, the NLO uncertainty reduces to $\sim 15\%$. For $pp$, the PDF uncertainties are below 4\% (not shown in \cref{fig:5020spectra}). The results with CT18ANLO and nNNPDF3.0 are in good agreement, while TUJU21 exhibits a different trend, crossing the other two around $p_T^h \sim 50\,\text{GeV}$. For OO collisions, EPPS21 nPDF uncertainties are substantially larger, exceeding $5\%$, and are mostly consistent with nNNPDF3.0 results, while TUJU21 again shows a distinct momentum dependence.

Finally, sizeable discrepancies arise between BKK, NNFF1.0h, and NPC23 FFs, which are not covered by the uncertainty bands of NNFF1.0h and NPC23. Predictions with NNFF1.0h lie about 25\% below those with BKK, while the NPC23-to-BKK ratio exhibits a pronounced momentum dependence. Nevertheless, all FF sets remain mutually consistent once the NLO scale uncertainty is taken into account.

\section{Conclusions\label{sec:discussion}}

In July 2025 the LHC collided oxygen ions for the first time, providing a unique opportunity to search for an unambiguous signal of jet quenching in a system with, on average, only about ten participating nucleons. Because the expected signal is small, it is essential to quantify precisely the difference in the initial perturbative production of high-momentum partons between oxygen–oxygen and proton–proton collisions. To this end, I have carried out a systematic study of the no-quenching baseline for the single-inclusive hadron nuclear modification factor, $R^h_\text{AA}$, in minimum-bias oxygen–oxygen collisions at $\sqrt{s_\text{NN}}=5.36\,\text{TeV}$. Using state-of-the-art nuclear parton distribution functions (nPDFs) and fragmentation functions (FFs), I have provided a robust estimate of the theoretical uncertainties associated with this baseline.

In the momentum window $20\,\text{GeV} \lesssim p_T \lesssim 70\,\text{GeV}$, the baseline can be controlled at the $\sim 5\%$ level, with nPDFs representing the dominant source of uncertainty. At lower and higher $p_T$, however, the combined nPDF and FF uncertainties become sizeable, potentially limiting the ability to disentangle medium effects from the baseline. The difference between neutral-pion and charged-hadron $R^h_\text{AA}$ is negligible compared with the nPDF uncertainties, indicating that both observables are equally suitable for probing medium-induced energy loss.

The largest systematic uncertainty in charged-hadron $R^h_\text{AA}$, particularly at low $p_T$, stems from the limited knowledge of oxygen nPDFs. The large $p$O dataset recently collected at the LHC will be crucial for constraining oxygen nPDFs in future global fits~\cite{Paakkinen:2021jjp}. At high $p_T$, I find non-negligible discrepancies between different FF extractions, which appear to originate from isospin-symmetry breaking in some sets. These discrepancies will hopefully be resolved in future global FF analyses, potentially aided by input from hadron spectra in $p$A collisions. In parallel, theoretical advances in charged hadron computations—NNLO accuracy in hadronic collisions~\cite{Czakon:2025yti} and N3LO in electron–positron annihilation~\cite{He:2025hin}—are paving the way towards substantially more precise FF determinations.

In conclusion, the charged-hadron nuclear modification factor in oxygen–oxygen collisions can be predicted to within $\sim 5\%$ in the most favourable kinematic range. Whether this precision is sufficient to observe a jet-quenching signal in such a small system will also depend on experimental uncertainties and the potential signal size~\cite{Huss:2020whe,Liu:2021izt,Zakharov:2021uza,Ke:2022gkq,Behera:2023oxe,Xie:2022fak,vanderSchee:2025hoe,Pablos2025}. In particular, a new source of experimental uncertainty arises in oxygen beams due to transmutation into other ion species~\cite{Nijs:2025qxm}. Finally, the collected oxygen–oxygen and neon–neon data at the same energy provide an additional handle: the ratio of hadron spectra between NeNe and OO collisions would largely cancel the nPDF uncertainties in the perturbative baseline, while still allowing for a potential medium-induced signal~\cite{vanderSchee:2025hoe}.

\begin{acknowledgments}

I thank Austin Baty, Jun Gao, Jannis Gebhard, Alexander Kalweit, ChongYang Liu, Emanuele Roberto Nocera, Patrik Novotny, Alexander Milov, Nicolas Strangmann, and Adam Takacs for helpful discussions and their interest in this work. I am particularly grateful to Jun Gao and ChongYang Liu of the NPC collaboration for providing the neutral-pion FFs. I also acknowledge the use of OpenAI’s ChatGPT to assist with language editing and improving the clarity of the manuscript.

This work was supported by the DFG through the Emmy Noether Programme (project number 496831614) and CRC 1225 ISOQUANT (project number 27381115).
\end{acknowledgments}

\appendix

\section{Rapidity dependence of $R^h_\text{AA}$\label{app:rapidity}}

\begin{figure}
    \centering
    \includegraphics[width=0.9\linewidth]{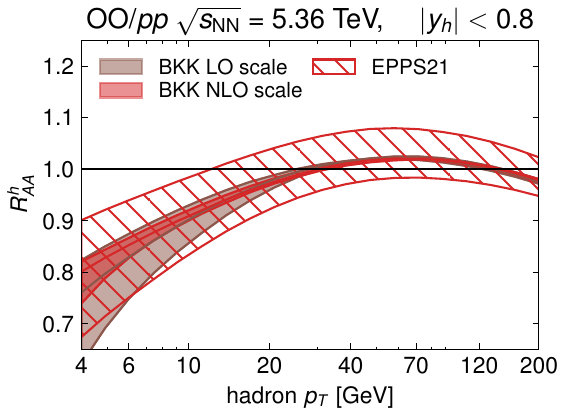}
    \includegraphics[width=0.9\linewidth]{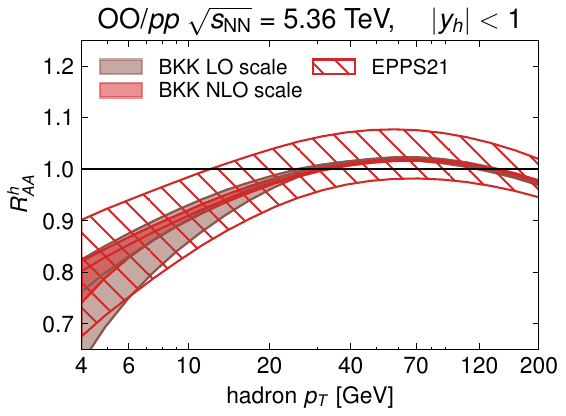}
    \includegraphics[width=0.9\linewidth]{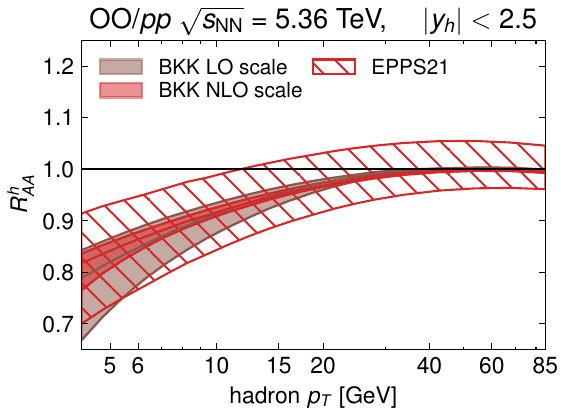}
    \caption{pQCD computation of charged hadron nuclear-modification factor baseline in oxygen-oxygen collisions at $\sqrt{s}=5.36\,\text{TeV}$ for (top) $|y|<0.8$, (middle) $|y|<1.0$ and (bottom) $|y|<2.5$. The colour bands indicate scale variation, while the hatched band represents 68\% nPDF confidence interval.}
    \label{fig:RAA2}
\end{figure}

In \cref{fig:RAA2}, I present the no-quenching baseline results for the single-inclusive charged-hadron nuclear modification factor $R^h_\text{AA}$ in oxygen–oxygen collisions at $\sqrt{s}=5.36\,\text{TeV}$ for different rapidity ranges, $|y|<0.8$, $|y|<1.0$, and $|y|<2.5$. These intervals correspond to the expected charged-hadron rapidity acceptance of the ALICE~\cite{ALICE:2018vuu}, CMS~\cite{CMS:2016xef}, and ATLAS~\cite{ATLAS:2022kqu} detectors. The results for $|y|<0.8$, already shown in \cref{fig:RAA}, are repeated here to facilitate comparison. For $|y|<2.5$, results are displayed only up to $p_T<85\,\text{GeV}$, matching the anticipated experimental momentum coverage.

The brown and red bands indicate the LO and NLO scale uncertainties, computed with EPPS21 nPDFs and BKK FFs, as in \cref{fig:RAA}. The red hatched band denotes the 68\% confidence interval for EPPS21. The results for $|y|<0.8$ and $|y|<1.0$ are nearly indistinguishable, while for $|y|<2.5$ the central value of $R^h_\text{AA}$ lies closer to unity in the range $40\,\text{GeV} \lesssim p_T \lesssim 85\,\text{GeV}$.
 
\section{Isospin symmetry breaking in NNFF1.1h fragmentation functions\label{app:isospin}}

In this section, I discuss the origin of the surprising suppression of $R_\text{AA}^h$ obtained with NNFF1.1h FFs in \cref{fig:RAA}.

\begin{figure}
    \centering
    \includegraphics[width=\linewidth]{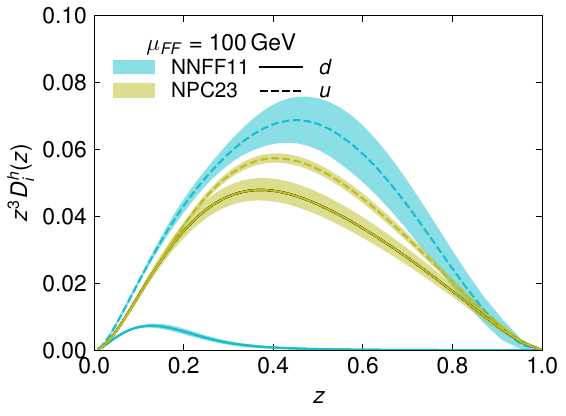}
    \caption{Fragmentation functions for down (solid lines) and up (dashed lines) to charged hadrons for different FF families. Colour bands show 68\% confidence intervals.}
    \label{fig:D}
\end{figure}

In \cref{fig:D}, I show the charged-hadron fragmentation functions $D^h_i(z)$ for down (solid lines) and up (dashed lines) quarks from NNFF1.1h and NPC23 at $\mu_{FF}=100\,\text{GeV}$. The vertical axis is scaled by $z^3$, motivated by the LO cross-section formula~\cite{Huss:2020whe}:
\begin{equation}
\frac{d\sigma^h}{d p_T^h} = \int_0^1 \frac{dz}{z}\left.\frac{d\sigma^i}{d q_T}\right|_{q_T=p_T^h/z} \!\times D^h_i(z,Q=p^h_T).
\end{equation}
For a steeply falling quark spectrum, $\frac{d\sigma^i}{dq_T}\sim q_T^{-4}$, the hadron cross section is proportional to the third moment of the FFs, i.e., the area under the $z^3D^h_i(z)$ curve.

For NPC23, the up- and down-quark FFs are very similar (\cref{fig:D}). This reflects the fact that NPC23 charged-hadron FFs are dominated by pions, which exhibit a high degree of isospin symmetry~\cite{Gao:2024dbv}. In contrast, NNFF1.1h shows a much smaller down-quark FF compared with the up-quark FF. The flavour separation of quark FFs is poorly constrained by data and typically requires additional assumptions. In NNFF1.1h~\cite{Bertone:2018ecm}, flavour separation was introduced following the procedure for kaons described in Appendix A of Ref.~\cite{Bertone:2017tyb}. Since charged kaons are not symmetric under $d\leftrightarrow u$ exchange, this approach naturally leads to FFs that violate isospin symmetry.

\begin{figure}[th]
    \centering
    \includegraphics[width=\linewidth]{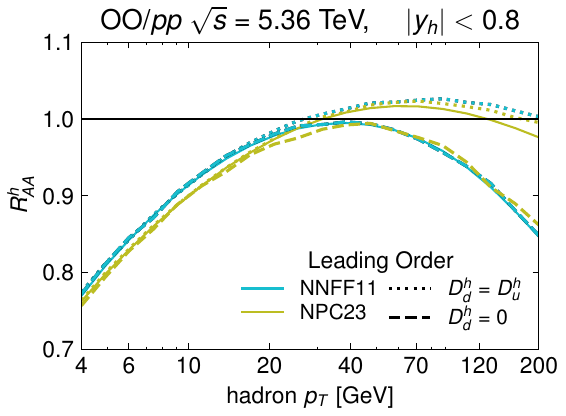}
    \caption{Leading order charged hadron $R_\text{AA}^h$ in  $\sqrt{s_\text{NN}}=5.36\,\text{TeV}$ oxygen-oxygen collisions. Lines show the central results for different FFs and EPPS21 nPDFs. Dashed lines are results with down quark FF set to zero, while dotted lines show results for strict isospin symmetry, i.e., $D_u^h=D_d^h$.  }
    \label{fig:FFiso}
\end{figure}

In \cref{fig:FFiso}, I demonstrate that this up–down difference drives the discrepancy observed in \cref{fig:RAA}. To this end, I performed LO cross-section calculations with modified FFs. The solid lines show $R_\text{AA}^h$ with the original NNFF1.1h and NPC23 FFs combined with EPPS21 nPDFs. The dotted lines correspond to computations where the down-quark FF is set equal to the up-quark FF (enforcing strict isospin symmetry). In this case, the NPC23 results remain essentially unchanged and consistent with the observed isospin symmetry in \cref{fig:D}. In contrast, the NNFF1.1h results become nearly identical to those of NPC23. Conversely, setting the down-quark FF to zero (maximal isospin asymmetry) leaves NNFF1.1h largely unaffected, while NPC23 develops the anomalous suppression of $R_\text{AA}^h$ at high momentum. These modifications only affect the high-$p_T$ region, since at $p_T<50\,\text{GeV}$ hadron production is dominated by gluon fragmentation~\cite{Huss:2020whe}.

\bibliography{citations}

\end{document}